# Muscle Synergy Patterns During Running: Coordinative Mechanisms From a Neuromechanical Perspective

MA Ye[1], LIN Shixin[1], FU Shengxing[2], LIU Yuwei[3], GUO Chenyi[4], LIU Dongwei[5], HOU Meijin[6,7]

(1. Research Academy of Grand Health, Faculty of Sports Sciences, Ningbo University, Ningbo 315211, China; 2. School of Sport Science, Beijing Sport University, Beijing 100084, China; 3. Auckland Bioengineer Institute, University of Auckland, Auckland 1010, New Zealand; 4. Department of Electronic Engineering, Tsinghua University, Beijing 100084, China; 5. School of Information Technology and Artificial Intelligence, Zhejiang University of Finance and Economics, Hangzhou 310018, China; National Joint Engineering Research Centre of Rehabilitation Medicine Technology, Fujian University of Traditional Chinese Medicine, Fuzhou 350122, China; 7. Key Laboratory of Orthopaedics and Traumatology of Traditional Chinese Medicine and Rehabilitation, Fujian University of Traditional Chinese Medicine, Fuzhou 350122, China)

Abstract: Running is a fundamental form of human locomotion and a key task for evaluating neuromuscular control and lower-limb coordination. In recent years, muscle synergy analysis based on surface electromyography (sEMG) has become an important approach in this area. This review focuses on muscle synergies during running, outlining core neural control theories and biomechanical optimization hypotheses, summarizing commonly used decomposition methods (e.g., PCA, ICA, FA, NMF) and emerging autoencoder-based approaches. We synthesize findings on the development and evolution of running-related synergies across the lifespan, examine how running surface, speed, foot-strike pattern, fatigue, and performance level modulate synergy patterns, and describe characteristic alterations in populations with knee osteoarthritis, patellofemoral pain, and stroke. Current evidence suggests that the number and basic structure of lower-limb synergies during running are relatively stable, whereas spatial muscle weightings and motor primitives are highly plastic and sensitive to task demands, fatigue, and pathology. However, substantial methodological variability remains in EMG channel selection, preprocessing pipelines, and decomposition algorithms, and direct neurophysiological validation and translational application are still limited. Future work should prioritize standardized processing protocols, integration of multi-source neuromusculoskeletal data, nonlinear modeling, and longitudinal intervention studies to better exploit muscle synergy analysis in sports biomechanics, athletic training, and rehabilitation medicine.

通信作者：刘东威，讲师，Email: dongwei.liu@zufe.edu.cn; 侯美金，讲师，Email: meijinhou@fjtcm.edu.cn

# 跑步过程中的肌肉协同模式：
# 神经生物力学视角下的协调机制

马晔[1]，林时鑫[1]，傅升星[2]，刘雨薇[3]，郭辰仪[4]，刘东威[5]，侯美金[6,7]

(1. 宁波大学体育学院大健康研究院; 2. 北京体育大学运动人体科学学院; 3. 奥克兰大学生物医学工程研究院；4. 清华大学电子系；5. 浙江财经大学信息技术与人工智能学院；6. 福建中医药大学康复产业研究院；7. 中医骨伤及运动康复教育部重点实验室)

摘要：跑步是人类最重要的运动形式之一，也是评估人体神经肌肉控制与下肢协调能力的典型任务。近年来基于表面肌电的肌肉协同分析逐渐成为热点。本综述围绕跑步情境下的肌肉协同，从协同的神经控制理论与生物力学优化假说出发，梳理协同分解算法及自编码器等新兴方法，综述不同年龄段跑步肌肉协同的发育与演进，总结跑步介质、速度、着地方式、疲劳与竞技水平等因素对协同模式的影响，并探讨了膝骨性关节炎、髌骨疼痛综合征、脑卒中等病理人群的协同特征。综合现有证据，下肢跑步协同在数量和基本结构上相对稳定，而其空间肌肉权重与时间募集模式对任务需求、疲劳及病理状态敏感，表现出较高的可塑性。当前研究在肌电通道布置、预处理流程和分解算法等方面仍存在方法学特异性，神经生理机制验证及应用转化证据亦相对不足。未来需要在处理流程标准化、多源神经肌骨数据融合及联合研究、非线性建模及纵向干预研究等方面进一步推进，以更好发挥肌肉协同分析在运动生物力学、运动训练及康复医学中的应用潜力。

中枢神经系统（central nervous system, CNS）在应对复杂运动控制任务时，采用了一种高效的协调策略，肌肉协同控制（muscle synergy control）[1]，从而减轻运动与姿势控制中所面临的计算负担。该机制通过将多个具有功能相关性的骨骼肌整合为“协同单元”，由统一的神经指令共同驱动，从而显著降低了神经系统在动作组织与调控过程中的自由度负担。

跑步是人类最重要的运动形式之一，可由少量基本的肌肉激活模式通过线性组合产生[2]。肌肉协同模式被用来探讨神经系统如何用简化的方式调控下肢与躯干的高自由度运动，从而应对不同的着地方式[3]、速度[2, 4]、地形[5]、疲劳程度[6, 7]或者跑步方式（平地或跑步机）[2, 8]等变量的影响。同时，神经肌肉骨骼系统损伤[9-12]及衰老[13]等会导致肌肉协同模式异常。

跑步中的肌肉协同研究可以用于研究肌肉控制策略，评估运动损伤风险，辅助制定康复干预方案。同时，还可以用于提升跑步效率与训练适应性等运动训练场景，具有很强的科学和应用价值。本综述系统性梳理了肌肉协同的理论基础与研究方法，并总结了健康人跑步及神经肌肉骨骼系统损伤人群步行中下肢肌肉协同模式的相关研究。本综述还评述了当前研究中的局限性与方法学不一致问题，并展望了未来研究方向，为生物力

学领域的专业读者提供参考。

## 1. 肌肉协同的神经起源与生物力学优化机制假说

肌肉协同究竟源自中央神经系统（central nerve system, CNS）的固有神经结构（即先天或通过学习形成的神经模块），还是只是生物力学结构或运动优化原则下的自然产物尚无定论。

神经起源假说认为CNS（如脊髓或脑干）中存在特定的神经回路，可同步驱动多个肌肉群，从而减轻中枢对高自由度肌肉骨骼系统的控制负担[14, 15]。该理论最早可追溯到Bernstein在1967年的研究，肌肉协同被定义为通过减少大脑控制的变量数量实现的肌肉简化控制[16]。多项动物（包括青蛙、大鼠以及猫等）实验研究通过微刺激、N-甲基-D-天冬氨酸（NMDA）离子注入及皮肤刺激等多种方法验证脊髓电路（spinal cord circuitry）的模块化组织结构[15]。

在脊髓蛙和大鼠中，当刺激电极沿脊髓腰段的头尾方向（rostro-caudal）和中外侧方向（medio-lateral）移动时，不同组的力场被激活。Mussa-Ivaldi 等人进一步发现，同时刺激两个脊髓部位会导致各个部位单独产生的力以矢量方式相加。在大多数情况下（83%），同时刺激两个脊髓部位所产生的力场，与分别刺激这两个部位所得力场的线性叠加结果基本一致[17]。后续的模拟研究进一步证实，能够生成稳定力场的青蛙后肢肌肉组合，与脊髓微刺激实验中观察到的共激活肌群高度相似[18]。这些研究结果共同支持一个观点：实验中观测到的力场是由以单元形式被激活的离散肌群所生成的，即所谓的肌肉协同，通过对这些协同的线性组合，可以构建出丰富多样的运动与姿势模式。

神经损伤后协同结构的变化亦为该假说提供了间接证据。例如，脑卒中后单侧偏瘫患者表现出足背屈模块融合，导致协同数目减少，运动复杂性降低，并与步态功能损害呈显著相关[19]。卒中后膝关节僵直步态人群也表现出简化的肌肉协同结构和肌肉激活的改变[20]。同时，卒中后功能损伤严重程度不同及病程不同情况下，脑卒中患者还表现为“保留”、“合并”与“分化”三种截然不同的肌肉协调模式[21]。

也有研究发现脑部运动指令与肌肉协同有较大的相关性。Mahdie Khaliq fard 等人采集了执行四种上肢功能运动（包括喝水、接电话、将水从水瓶倒入水杯中以及）任务时的脑电和肌电信号，使用Elman递归神经网络对脑电中提取的运动指令及肌电中提取的肌肉协同之间的非线性映射关系进行了解码建模，并通过十折交叉验证发现运动指令与肌肉协同间的相关系数约为85%[22]。

另一种观点强调，肌肉协同可能非源自中枢神经系统内部的固定神经模块，而是由于肢体生物力学特性及常见的运动优化准则自然生成的结果。Kutch 与 Valero-Cuevas 通过仿真研究与尸体实验指出，肢体运动受制于物理规律与简单控制策略（如肌肉抵抗被

拉伸的反射式机制），可自动生成低维度的肌电模式，呈现出类似“协同”的结构[23]。他们进一步指出，肌肉骨骼的几何约束会限制可能的激活空间，因此，即使 CNS 能够独立控制每块肌肉，最终的输出仍可能因生物力学耦合而呈现出类似协同的激活组合。由此可见，仅凭表面肌电中观测到的低维结构并不能直接证明协同具有神经起源。其他计算建模研究亦支持该假说。例如，在以能耗最小化或力学效率最大化为目标的优化框架中，模型也会自然生成重复性肌肉激活组合，表现为类似协同的低维结构。例如，一项研究将预定义的协同结构引入步态静态优化模型，并与传统优化方法进行比较，结果显示引入协同结构的模型在预测肌肉激活方面并未显著优于传统不包含协同结构的优化模型[24]。这些发现表明，部分协同结构可能是由任务需求与生物力学约束共同塑造，而非完全由神经系统固有结构决定。

综合上述两类观点，我们认为肌肉协同可能具有明确的神经生理基础，同时其结构受到个体特有的肌骨力学约束及其他生物力学属性的影响。CNS 在组织运动控制时，可能并非单纯依赖神经模块或生物力学约束，而是综合利用二者所提供的信息，并在任务需求、能量最优及运动经验等因素的共同作用下，通过学习与神经可塑性逐渐强化，从而形成稳定的“协同化”运动控制单元。

## 2. 基于表面肌电信号的肌肉协同模式提取方法

肌肉协同分析所依赖的数据基础是经过预处理、归一化与时间对齐后的“肌肉×时间点”肌肉激活矩阵。首先，需要在采集阶段准确定位各肌肉的肌电极放置部位，严格遵循标准化备皮流程，以获取高质量的表面肌电信号，并最大限度减少运动过程中因皮肤-电极相对位移等因素导致的运动伪影。目前跑步协同研究主要采用双极表面肌电电极，采集下肢 8 到 24 块主要肌群的信号[3, 7]。然而需要注意的是，若某块在协同中实际发挥作用的肌肉未被记录，可能导致协同被人为拆分，或使其他肌肉权重偏高[25]。因此，建议在设计实验时进行尽可能全面的肌群覆盖。

表面肌电信号通常需要经过带通滤波（20~450 Hz）、整流以及采用截止频率为 5~20 Hz 的低通滤波获得包络线。随后需进行幅值归一化和时间标准化。幅值归一化通常使用各通道跨试次的最大值或最大自主收缩，将不同肌肉的激活幅度统一至 0~1 区间，避免因某些高幅值肌肉错误的主导矩阵分解结果。时间标准化则是依据步态周期（例如从同侧足跟触地至下一次触地）对数据进行切分，并将每周期内的数据点插值至统一的长度（如 100 或 200 个时间点）。这一过程有助于在统一的步态相位框架下进行跨试次（inter-trial）及跨个体（inter-subject）分析及比较。

表面肌电信号的预处理过程会对最终肌肉协同结构产生一定影响。一般而言，滤波截止频率或归一化方式等参数的细微差异不会凭空生成新的协同结构（例如使用截止频

率为 6~10 Hz 的低通滤波器提取包络线并不会显著改变协同结构和激活波形[26]，但可能影响协同数量[26]。因此，为了提高不同研究之间的可比性，有必要在协同分析的信号处理流程方面进一步推动方法学标准化。

## 2.1 用于提取肌肉协同的矩阵分解技术

当前生物力学领域用于提取肌肉协同模式的主要方法为矩阵分解技术，包括主成分分析（principal component analysis, PCA）、独立成分分析（independent component analysis, ICA）、因子分析（factor analysis, FA）以及非负矩阵分解算法（non-negative matrix factorization, NMF）等。

主成分分析（PCA）通过正交线性变换将高维数据投影到低维线性子空间，以最大化投影方差并实现降维。对中心化数据矩阵，主轴（主方向）由数据协方差矩阵（或对标准化数据使用的相关矩阵）的特征向量构成，并按对应特征值（方差贡献）从大到小排序；数据在这些主轴上的投影称为主成分，主成分之间互不相关。在数值实现上，PCA 常通过奇异值分解（SVD）获得主轴与得分，这与对协方差矩阵的特征分解在数学上等价。PCA 不保证统计独立性，也不施加非负约束，在肌电信号分析中（尤其是肌肉激活模式时）可能降低其生理可解释性，故常与强调非负性的 NMF 或强调独立性的 ICA 进行比较与权衡[27]。

独立成分分析（ICA）旨在从线性混合观测中分离出统计独立的潜在源信号，其可辨识性通常要求源信号为非高斯（至多一个高斯源）。在估计过程中，ICA 通过最小化分量间的互信息或最大化非高斯性来求解独立分量，常用的对比函数包括基于峰度的准则、基于近似负熵（negentropy）的准则、最大似然或最大化信息传递 (Infomax) 准则等[28]。为去除二阶相关并简化求解，ICA 在计算前通常对数据进行预白化，使协方差矩阵接近为单位阵。在算法实现方面，Infomax ICA 可视为通过最大化输出熵来提取独立分量，其等价于在给定非线性模型下最小化互信息或实现最大似然估计；Kernel ICA 则借助核方法在再生核希尔伯特空间中度量并最小化高维特征空间中的依赖性。ICA 已广泛用于肌电信号分离、伪迹去除与神经解码等任务。在肌电协同分析中，ICA 得到的基向量（或混合矩阵的列）与独立分量共同描述肌群活动的独立驱动模式。该算法相比 PCA 更强调统计独立性。但其同样不施加非负约束，在生理可解释性方面同样存在瑕疵。因此，ICA 在协同分析中的使用应与研究目的结合，在统计独立性与生理可解释性之间进行权衡[29]。

与主成分分析（PCA）不同，因子分析（FA）基于共同因子模型$\Sigma = \Lambda\Lambda^{T} + \Psi$将变量的协方差分解为公共方差与独特方差两部分。FA 的因子载荷矩阵$\Lambda$通常通过最大似然（ML）、广义最小二乘（GLS）或主轴因子法（PAF）等方法估计，并常配合正交或

斜交旋转以提升可解释性；其求解并非一般地等同于对协方差矩阵做特征值分解。因子个数不宜仅以“特征值>1”的 Kaiser 准则判定，该准则主要针对 PCA 的相关矩阵提出，属启发式方法，且易产生偏高估计。在 FA 中，更推荐结合平行分析、碎石图、MAP 统计量、似然比检验、信息准则及残差拟合等方法综合决定因子数。FA 可能产生正负载荷且受旋转影响，生理解释同样需要谨慎对待[30]。

非负矩阵分解（NMF）通过将非负数据矩阵 $V$ 分解为基矩阵与系数矩阵的乘积$V \approx WH(W,H \geq 0)$，以加性、部件化的方式表示数据[31]。常用的目标包括 Frobenius 范数、Kullback-Leibler 散度与 Itakura–Saito 散度，分别对应高斯、泊松及乘性噪声等不同的数据生成假设。主流求解策略有乘法更新、投影梯度/坐标下降（如 HALS）以及交替非负最小二乘（ANLS/NNLS）。与以最大化方差解释率或正交特征向量为目标的 PCA 不同，NMF 的核心在于非负约束，使其分解结果更符合表面积淀信号的生理特性。具体而言，NMF 分解得到的基矩阵和系数矩阵可以对应到肌肉协同中的空间模块及时间激活模块。在实际协同建模应用中，常结合稀疏、平滑等正则化手段以提升稳健性与生理可解释性。

与 PCA、ICA 和 FA 等算法相比，NMF 在肌肉协同提取中的主要优势体现在可解释性与建模灵活性：（1）神经元的放电率始终为非负，且突触强度通常不改变符号，这与 NMF 在建模过程中对基础向量和系数同时施加非负约束的特点高度契合，而 PCA、ICA 和 FA 等算法并不具备该约束条件。（2）NMF 不要求基向量之间具备正交性或统计独立性，对数据分布的依赖性较小，具有更强的适应性。（3）在动态运动任务中，NMF 表现出更优的性能，具体体现在其提取的协同模式更具稀疏性、一致性更好并且方差解释度更高[26]。

表 1：肌电协同模式主要提取算法

Table 1: Major algorithms for EMG synergy pattern extraction

| 方法 | 主假设 | 输出结构 | 优点 | 局限性 |
|---|---|---|---|---|
| PCA | 正交、最大方差 | 正负载荷的正交模式 | 稳定、计算快、用于降噪与探索性分析 | 不符合 EMG 非负性；难以解释协同结构 |
| ICA | 独立性、非高斯性 | 统计独立的模式 | 可分离驱动源、用于伪迹处理 | 无非负约束，解释性弱；依赖预白化 |
| FA | 协方差由公共因子驱动 | 旋转后的因子载荷 | 能表达协方差结构 | 载荷可正负；受旋转影响；解释需谨慎 |
| NMF | 非负性、加性结构 | 生理合理的非负模块 | 最符合肌肉激活特性；解释性强 | 解不唯一；排序分解的协同向量；对噪声敏感 |

综上所述，NMF 因其良好的生理可解释性与建模适应性，被广泛认为是在肌肉协

同提取中性能最优的算法。目前已有研究统计显示，约 70%的肌肉协同文献使用 NMF 作为主要方法[26]。

## 2.2 肌肉协同的向量表示形式

使用 NMF 方法能够提取两种肌肉协同类型：空间固定协同 (spatial fixed, SF) 和时间固定协同 (temporally fixed，TF)[32]。一般来讲，NMF 算法将原始的表面肌电矩阵 $\boldsymbol{E}$ 分解为空间肌肉权重矩阵 $\boldsymbol{W}$ 与时间募集模式矩阵 $\boldsymbol{C}$。算法起始时对 $\boldsymbol{W}$ 和 $\boldsymbol{C}$ 赋予非负随机值，随后通过迭代更新，使实际矩阵 $\boldsymbol{E}$ 与重构矩阵 $\boldsymbol{E}^*$之间的平方误差之和$e$最小。其基本形式可以表示如下：

$$\mathrm{E} = \mathrm{WC} + e \tag{1}$$

$$e = \sum_i \sum_j \left(E_{i,j} - E_{i,j}^*\right)^2 \tag{2}$$

当肌肉协同数量 $N_{\mathrm{syn}}$ 确定后，某肌肉的肌肉活动 $\boldsymbol{M}_i$ 可以使用肌肉权重 $\boldsymbol{W}_i$ 与时间募集模式矩阵 $\boldsymbol{C}$ 按照如下公式重建：

$$\boldsymbol{M}_i = \sum_{j=1}^{N_{\mathrm{syn}}} w_{i,j}\ C_j \tag{3}$$

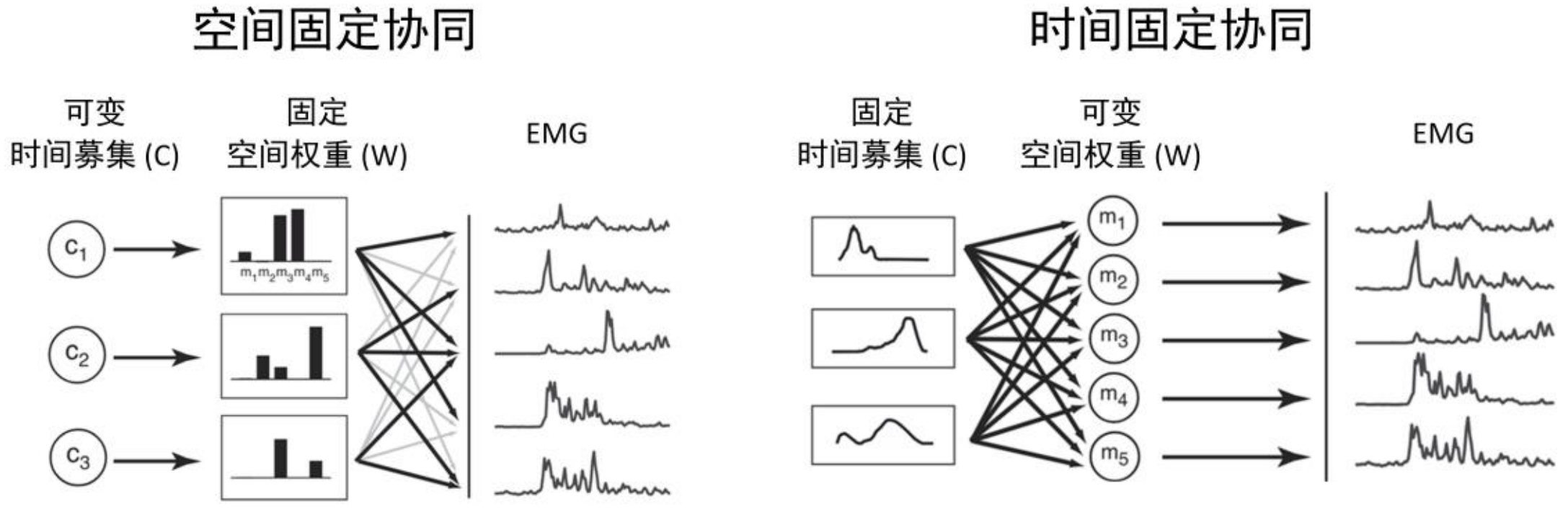

图 1：空间固定及时间固定协同示意图

Fig. 1: Schematic illustration of spatially fixed and temporally fixed synergies

如图 1 所示对于空间固定协同，空间肌肉权重矩阵 $\boldsymbol{W}$ 保持不变，但其时间募集模式 $\boldsymbol{C}$ 可在每个时刻、每次实验中发生变化。相反，对于时间固定肌肉协同，时间募集模式 $\boldsymbol{C}$ 保持不变，而其空间肌肉权重 $\boldsymbol{W}$ 可随实验条件而改变[32]。

肌肉协同的分解方式应根据所研究的运动模式（例如步行等节律性运动、全身伸取等离散动作[33]、或平衡扰动等姿势控制任务[1]）以及研究所关注的核心问题进行选择。对于跑步任务，多数研究采用空间固定协同模式，以提取相对稳定的肌肉协同结构组成，并同时分析其在跑步周期中的时间激活特性[34]。使用时间固定协同模式，可着重分析跑步中肌肉电活动爆发的时间结构（例如支撑相与摆动相的“时间原型”）[35, 36]。此外，当研究需要同时兼顾及比较空间与时间两类组织形式，或检验他们的分类能力与表征效果

时，还可以采用基于样本的时-空联合分解方法（space-by-time decomposition），如非负矩阵三分解等[33]。

### 2.3 非负矩阵分解技术的注意要点

尽管 NMF 算法在生理可解释性方面具有明显优势，但其分解结果不唯一，并且对初始化非常敏感。在实际应用中首先需要确定适当的协同数量。假设采集的下肢肌群数量为 $n$，通常将协同模式的数量 $k$ 在 1 至 $n$ 的范围内进行迭代计算，并计算各情形下的方差解释度（Variance Accounted For, VAF）。多数研究以能够使肌电信号总 VAF 达到 90%或者 95%的最小协同数作为最终选定的 $k$ 值。为确保各肌肉的重构质量，一些研究进一步采用“局部 VAF”指标，即要求每个肌肉通道的 VAF 均超过某一临界值（如 0.8），以判断是否需引入额外协同来解释某些特异性肌肉的激活模式。此外，还可结合误差曲线的“拐点法”辅助确定协同数量：例如若新增一个协同所带来的 VAF 提升不超过 1%，则可认为无需继续增加协同数量。在群体研究中，为增强不同受试者间协同结构的可比性，通常采用“最小协同数原则”，即选取能在所有受试者中实现高 VAF 重构的最小 $k$ 值，从而提升跨个体分析的解释一致性与生理合理性。

VAF 指标用于衡量重建信号对原始信号的拟合程度，可采用非中心化皮尔逊相关系数形式计算。为了不同试次、场景和个体等情况可以比较及统计，通常可根据具体的 VAF 计算结果选取一个具备普遍性的肌肉协同个数。

$$\mathrm{VAF} = 1 - \frac{\sum_{i=1}^{N_{\mathrm{syn}}}\sum_{j=1}^{n}\left(e_{ij}\right)^2}{\sum_{i=1}^{N_{\mathrm{syn}}}\sum_{j=1}^{n}\left(E_{ij}\right)^2} \tag{4}$$

NMF 属于典型的非凸优化问题，而其求解方法是梯度下降迭代优化，属于贪心算法，易陷入局部最优，且初始参数的设定会显著影响分解的结果。为提高分解结果的可靠性，在协同数量确定后，通常需要在 $n$ 个不同初始状态下运行迭代算法 $i$ 次，并从中选取重构误差最小或方差解释度（VAF）最高的解作为最协同模式。其中参数 $i$ 不充分会导致求解不收敛，而 $n$ 过小则可能导致找到的解质量较低。在下肢肌肉协同的提取领域，不同研究对重复迭代次数的设定差异较大。例如，有研究采用 20 次迭代[13, 37]，也有研究设置为 30 次[38]，部分研究甚至高达 3000 次[3]，但较少研究明确表述其初始条件设定细节。

需要特别注意的是，NMF 算法在不同试次中得到的肌肉协同模块往往不存在固定的排序。因此，为了能够从多个试次乃至多个受试者中探究肌肉协同模式，必须把协同向量的顺序统一排列。常用的协同排序与匹配方法包括以下几种形式：（1）根据经验选取一组参考协同向量[35]，然后分别计算其他试次中的每个协同与该参考向量之间的相似度（如余弦相似度[2, 35]）。若相似度大于某一阈值（例如 0.6），则认为该协同属于同一个

协同[2, 35]；（2）将所有试次获得的协同向量作为样本，利用聚类算法（例如 *k*-means 等）进行分组[7, 37, 38]。若同一试次内的多个协同向量被分到了同一组，则将更接近簇中心的向量分配到这一组。此外，还可通过评估协同结构在不同试次、不同初始化条件、不同受试者或不同任务条件下的重现程度来量化协同的稳健性，以判断协同是否具有生理一致性，而非由噪声或算法不稳定性引起[35, 39]。

基于表面肌电信号的矩阵分解算法本质上属于对表面肌电数据的统计建模技术，其所得协同结构仅反映肌电信号在数学意义下的低维模式。因此，尽管这些结果在一定程度上解释部分运动控制现象，但其本身并不构成严格的神经控制模型。当前文献中由于数学方法选择不统一、研究设计存在差异以及协同结果对预处理及计算细节高度敏感，研究者在解读肌肉协同模式时需保持谨慎态度，应基于系统系统性综述和多项证据进行分析，而不可对单一研究结果作过度推断。

在协同数量的确定上也应审慎权衡。研究者需结合既有文献、运动控制理论及协同结构的胜利合理性进行判断。当两个协同的时间募集模式及空间肌肉权重均高度相似时，应重新审视协同数量的设定及结果的生理可信度。此外，施加于协同分解算法的优化目标函数也需谨慎选择。原则上，所采用的优化准则决定了最终获得的协同结构，不恰当的目标函数可能将问题简化为一般性的拟合问题，而偏离对神经控制本质的探讨。因此，研究中应明确所引入的正则项或目标函数是否反应了运动控制的固有特征，以避免对协同结构作出过度或错误的生理解释。

### 2.4 其它新兴方法

基于全连通神经网络构建的标准编码器（encoder）-瓶颈层（bottleneck）-解码器（decoder）的自解码器（Autoencoder）架构，可用来提取 sEMG 数据中的肌肉协同模式[40]。在该框架中，编码器部分提取的潜在特征可表征为肌肉协同的时间特征（即前述的“时间原型”），而解码部分对各个肌肉的重构权重则被视为肌肉协同的空间特征[40]。

与传统的线性矩阵分解方法（如 PCA，ICA，NMF）相比，自解码器架构的神经网络在协同提取方面有以下优势：（1）能够在单一协同中表示拮抗肌成对关系（而 NMF 的非负性约束无法做到）；（2）在 EMG 重构任务中可获得更高的方差解释度。换言之，自编码器提取的协同模式与数据的拟合显著更好，并且保留了生理上有意义的负相关，而这些信息在传统线性方法中常被忽略[41]。

## 3. 健康人跑步中的协同模式及其可塑性

在前述讨论中，我们假设肌肉协同是一种相对稳定的模块化结构，即其在较长时间尺度上保持基本一致。然而，人体神经-肌肉-骨骼系统具有高度的可塑性，并会在个体发育、衰老、运动训练、技能学习过程中持续发生结构与功能变化。因此，为在不同生

理状态下机体仍能成功完成目标运动任务，机体的肌肉协同结构需要在运动发育过程中发生重塑，或在运动适应、新运动技能学习以及损伤后功能恢复等阶段进行一定程度的调节甚至重构[42]。

### 3.1 不同年龄阶段跑步肌肉协同的发育与演进

已有研究表明，无训练背景或者缺乏跑步经验的普通成年人，在早期肌肉协同模式（3~6 岁的学龄前儿童）的基础上分化出了更多协同变体，并且表现出更大的协同稀疏性[37]。以 Cheung 等人的研究为例，他们采集了幼儿与无跑步训练经验成人的右侧躯干及下肢 15 块肌肉跑步中的表面肌电信号。结果显示，幼儿群体可提取 7 个协同模式，而普通成年人群体中呈现出 11 个协同模式。普通成人群体与幼儿共享 6 个协同，并且成人群体中高度碎片化的“仅胫前肌”及“阔筋膜张肌与臀大肌”协同是从幼儿的三个协同中分化而来[37]。

成年人的跑步协同分化现象可以认为是一种适应重组过程，即在神经-肌肉骨骼系统不断成熟的背景下，对早期协同进行调整与重构，以适应形态结构、力量控制和神经调节策略的变化[43]。学龄前儿童常表现出明显的共激活现象，例如踝关节跖屈肌与背屈肌的同时募集，这是对姿势与平衡控制能力尚未成熟的神经机制的一种补偿策略[44, 45]，以维持跑步中的稳定性。随着年龄增长和神经控制能力提升，共收缩需求降低，运动控制系统能够更加精细的调动特定肌群子集，并淘汰部分学前期协同模式，形成更具分化度与任务特异性的成人协同结构。总体而言，儿童协同向成人协同的转变提高了运动控制的灵活性和场景适应能力，但也可能以一定的执行效率为代价。

若仅基于单侧下肢的表面肌电信号，已有文献指出四个基本协同即可满足平地及跑步机跑步任务的控制需求[8, 34]。具体包括：（1）承重反应期协同：由膝伸肌、髋伸肌和外展肌主导，用以承接地面反作用力，稳定下肢负荷；（2）推进阶段协同：主要由踝关节跖屈肌贡献，在支撑末期产生推进力；（3）摆动早期协同：主要由足背屈肌群贡献；（4）摆动末期与落地准备期协同：用于控制表动末期的减速与落地准备，步行与跑步均表现为膝屈肌主导，而在跑步中足背屈肌的募集更为显著，用以实现脚部前摆后的姿态修正[8]。

老年人的肌肉协同整体上与青年人类似，但在部分关节协同上存在显著差异。例如，老年人在髋关节屈伸以及膝关节屈曲相关肌群（例如阔筋膜张肌、臀大肌及股二头肌）为主的肌肉协同模式上，与青年人相比表现出较大差异[13]；同时，其时间激活协同模式也呈现出更高的个体内变异性[13]。老龄化往往伴随肌骨系统多种生物力学特性的改变，这些改变对功能性步态施加了不同的力学约束，从而需要通过对肌肉协同进行重新调整以维持有效的运动控制[13]。在跑步中，老年人较青年人表现为髋关节屈曲活动度增大而

伸展活动度减小，并倾向于依赖更高的髋屈肌功率以补偿踝跖屈肌推蹬能力的不足[13]。此外，老龄化还会导致中枢神经系统运动相关脑区结构与神经化学方面的改变，以及脑功能连接网络的改变[46]，并引起脊髓运动回路及中间神经元的退行性变化[47]。因此，脑-脊髓层面的年龄相关改变（如网络重组、反射通路调整、神经元损伤等）与外周肌骨生物力学特性的改变相互作用，共同推动并塑造了老年人肌肉协同模式的重塑过程。

### 3.2 不同跑步条件下肌肉协同模式的变化

成年人跑步中的肌肉协同模式在不同跑步介质、速度、着地模式等条件下具有较高的稳定性。与平地跑步相比，跑步机上的协同空间权重基本一致，不受跑步环境差异的影响；但协同激活原型（motor primitives）整体呈现更高的规律性，其中用于推进的激活原型时间上出现前移[8]。跑步机所施加的空间与感觉约束可能迫使中枢神经系统采用不同于自由地面跑步的神经控制策略，从而在时间激活维度上表现出更加规整的神经控制策略。

跑步速度与坡度的变化并不会显著影响肌肉协同的结构，包括空间肌肉权重及时间募集模式均保持高度稳定[2]。足部着地模式（前足着地及后足着地）对跑步机跑步中的协同结构影响亦相对有限。Nishida 等人[3]在 5~15 Km/h 的跑步机跑步条件下采集了双侧下肢及躯干共 24 块肌群的表面肌电信号，并从中提取了 6 个协同模式。结果显示，不同足部着地模式下协同的空间结构基本一致。人体主要通过调整胫骨前肌的协同权重以及对应的激活模式（包括激活时相、激活持续时间及峰值幅度）来适应不同的足部触地模式[3]。

随着竞技水平的提升，成年人的肌肉协同模式出现“合并”现象。以 Chung 等人的研究为例，无训练经验的普通成人存在 11 个协同，新手跑者减少为 7 个，精英运动员仅保留 6 个协同模式[37]。协同的“合并”本质上是将原本分散的控制指令空间缩至更小、更高效的指令子空间，从而减少控制自由度与运动变异性，提高控制效率[37]。

现有证据表明，协同合并可通过两种主要机制改善跑步效率：(1)在跑步支撑末期，小腿三头肌、股四头肌、臀大肌与阔筋膜张肌的协同激活增强，可显著提升推进能力[43]。该模式与肯尼亚精英选手的激活策略高度一致[48]。(2）通过背阔肌与踝背屈肌的协同，使臂摆与腿摆实现功能性同步，使臂摆能抵消下肢角动量、稳定躯干，并在高速度下提供额外加速度，从而减少能量损失、改善跑步经济性[37]。

然而，部分协同的合并可能导致增加能量损失。例如，在跑步初始接触期的踝跖屈肌与背屈肌的共同收缩，会提高踝关节刚度[49]和垂直载荷率[50]，但也能够增加能量损失。在支撑早期及摆动早期躯干肌肉与臀大肌的同时激活可能引发不自然、低效率的臂摆而增加能量损失[51]。总的来说，跑步经济性的提升不仅依赖肌力、体能或技术，更关键在

于通过训练塑造高效的协同组织方式，即“合并哪些协同、如何同步”。在强化推进与臂—腿协调的同时，还需避免诱发提高关节刚度与冲击载荷的“低效协同合并”，以实现真正意义上的高效率跑步控制。

## 3.3 运动疲劳对跑步肌肉协同模式的影响

运动疲劳通常并不会改变协同的数量，但会显著影响协同的空间肌肉权重及时间募集模块[6, 38]。例如，在 400 米冲刺跑末段，髋关节肌群（如骨直肌、臀大肌、阔筋膜张肌）的协同权重明显增加，而小腿肌群权重保持不变或下降，可以解释为身体通过在“近端”增加输出来补偿“远端”肌群的疲劳[7]。跑步机疲劳干预研究同样显示，以股四头肌及小腿肌群为主导的协同中，疲劳后股二头肌及胫骨前肌的权重显著增加；以腘绳肌及胫骨前肌为主导的协同模式中，腓肠肌内侧头与胫骨前肌的权重亦有所上升；同时，股四头肌内侧头和外侧头、腓肠肌和比目鱼肌的权重呈下降趋势[6]。这些变化共同表明，疲劳状态下 CNS 会在维持协同数量不变的情况下，通过调整协同内部的肌肉构成来重新分配负荷，以维持关节控制与姿势稳定性。

除了空间权重外，协同的时间募集模块也会在疲劳后出现明显改变。Xu 等人通过疲劳干预研究发现摆动末期及承重反应期协同的激活持续时间在疲劳后显著缩短，激活的时相延迟；推进期的激活时长显著延长，摆动末期协同的峰值时刻显著延后[38]。这些结果表明，疲劳后通过改变激活时相、时长和峰值位置来影响步态周期中不同相位的肌肉募集结构。

疲劳状态表现出的协同激活时间窗的延长，即协同曲线“变宽”，反应出不同肌群间更高程度的共同激活。这通常被视为中枢神经系统在肌肉疲劳、关节稳定性下降时的一种补偿策略，通过增加某些协同激活的持续时间来增强整体的稳定性[32]。在不稳定条件下行走的研究中，当外界扰动增强时，肌肉协同的激活曲线明显变宽（即激活曲线的半高宽增加），这意味着相比正常行走，协同保持激活的时间更长，来应对外界的不确定性[4]。

将这一现象类比到长距离跑步的后程，当疲劳累积时，跑者可能会在无意识的提前激活某些肌群，并延长其激活时长，从而抵消肌肉输出下降或关节晃动增加的影响。例如，在支撑期，股四头肌与腘绳肌可能出现更长时间的同步激活，以提升膝关节的稳定性；小腿肌群可能在支撑期延长激活时间以维持踝关节的稳定。这种协同激活的虽有助于维持关节稳定性与步态一致性，但代价是能量消耗增加，并可能削弱原本清晰的协同分界[4]。

目前关于超长距离耐力跑过程中协同变化的研究较为有限，但可依据现有短时疲劳研究推测其可能的变化趋势：当疲劳仍处于可控范围时，中枢神经系统通常会在维持原

有协同结构的前提下，通过调节协同内部的权重分布来进行补偿；当特定肌群出现明显功能下降甚至接近“失能”时（如小腿肌群严重疲劳），身体可能被迫采用全新的协调模式。肌肉协同的变化主要反映神经肌肉层面的适应，而若疲劳来源以心肺系统为主（如通气不足或循环限制），肌肉本体功能尚可时，协同结构可能保持稳定，仅表现为激活幅度减弱。因此，对肌肉协同的分析尤其适用于揭示肌肉局部疲劳对运动协调的影响。

表 2：跑步中的肌肉协同研究

Table 2: Studies on muscle synergies during running

| 文献 | 受试者及采集通道 | 肌肉协同主要结果 |
| --- | --- | --- |
| Kakehata et al. [7] | 9 人，14 通道单侧躯干及下肢：RF, BF, VL, VM, Gmax, Gmed, TFL, TA, SOL, LG, RERO, RAS, EO, ESL3 | ● 7 个空间固定协同；<br>● 400 米冲刺跑后无显著变化：<br>■ 髋关节相关肌肉 (RF, TFL, Gmax) 权重显著变化；<br>■ 股直肌激活时相显著偏移；<br>■ TFL 和 Gmax 激活显著提升。 |
| Nishida et al. [3] | 10 人，24 通道双侧躯干及下肢：MG, LG, SOL, TA, VL, RF, BF, TFL, AL, Gmed, Gmax, ES | ● 6 个空间固定协同；<br>● 不同着地模式（RFS，FFS）下协同结构保持稳定；<br>● 5~15km/h 跑步机跑步速度变化不影响协同数量；<br>● TA 在着地前的权重有显著差异；<br>● 协同的激活模式（时相、时长、峰值）有显著变化。 |
| Santus et al. [34] | 135 人，13 通道单侧下肢：Gmed, Gmax, TFL, RF, VM, VL, ST, BF, TA, PL, MG, LG，SOL | ● 2.2~3.0 m/s 跑步机跑步协同模式稳定；<br>● 4 个空间协同:<br>■ 承重反应期（膝伸肌及臀肌）<br>■ 推进期（跖屈肌）<br>■ 摆动早期（足背屈肌）<br>■ 摆动末期（膝屈肌、足背屈肌） |
| Mileti et al. [8] | 30 人，13 通道单侧：Gmed, Gmax, FL, RF, VM, VL, ST, BF, TA, PL, GM, GL, SOL | ● 2.8 m/s 平地及跑步机跑步，1.4 m/s 平地及跑步机走路中协同稳定；<br>● 4 个空间固定协同:<br>■ 承重反应期（膝关节伸肌、髋关节伸肌和外展肌）<br>■ 推进阶段（踝关节跖屈肌）<br>■ 摆动早期（足背屈肌）<br>■ 摆动后期及落地准备阶段（膝关节屈肌、足背屈肌）<br>● 跑步介质不影响走路及跑步的主要协同模式。<br>● 跑步机走跑中的推进阶段激活曲线峰值前移。 |
| Saito et al. [2] | 8 人，10 通道单侧下肢：VI, VL, VM, RF, BF, SM, AM, AL, MG, TA | ● 三种速度（2.5, 3.3, and 4.1 m/s）和两个倾斜角度（0°, 5.71°）下跑步机跑步协同的空间肌肉权重及时间募集模式均基本稳定；<br>● 4 个空间固定协同：<br>■ 右侧脚跟着地（股四头肌）<br>■ 协同一之后（腓肠肌）<br>■ 支撑后期及摆动期（内收大肌、胫骨前肌）<br>■ 右侧脚跟着地前（长收肌、腘绳肌） |
| Xu et al. [38] | 12 人，9 通道单侧下肢：Gmax, BF, RF, VM, VL, TA, MG, LG, SOL | ● 疲劳前后 12 km/h 跑步机跑步：<br>■ 协同数量不变；<br>■ 协同模式（包括肌肉权重及时间募集模式）均有变化。<br>● 6 个空间固定协同：<br>■ 摆动末期及承重反应期（协同 3 和 6）：股四头肌、臀大肌<br>■ 推进阶段（协同 2 和 4）：腓肠肌、比目鱼肌<br>■ 摆动前期（协同 1）：胫骨前肌<br>■ 摆动末期（协同 5）：股二头肌 |

注释：RF: rectus femoris, BF: biceps femoris long head, VL: vastus lateralis, VM: vastus medialis, Gmax: gluteus maximus, Gmed: gluteus medius, TFL: tensor fasciae latae, TA: tibial anterior, SOL: soleus, LG: lateral gastrocnemius, MG: medial gastrocnemius, RERO: peroneus longus, RAS: rectus abdominis, EO: external oblique, ES: erector spinae, ESL3: erector spinae at L3, SM: semimembranosus, ST: semitendinosus, AM: adductor magnus, AL: adductor longus, RFS: rear-foot strike, 后足着地, FFS: fore-foot strike, 前足着地

综上，疲劳引起的协同变化再次说明，中枢神经系统更倾向于在既有协同框架内灵

活的时间与空间调节，而非重新建立全新的运动协调结构。换言之，协同提供了一套稳定但可塑的运动控制基础，使中枢神经系统能够根据任务需求、疲劳状态及外界扰动进行快速适应。

### 3.4 跑步中的时间固定协同模式

上述研究均基于空间固定协同模式。从时间固定协同模式的角度看，跑步同样呈现出稳定的模块化结构。Ivanenko 等人[36]与 Cappellini 等人[52]分别基于 12~16 块及 32 块单侧下肢及躯干肌群的表面肌电信号，均确认跑步中存在5个主要的时间固定协同模式。并且，这 5 个基础的时间激活模式在形态与时相上与步行的时间协同高度相类，其所对应的协同肌肉组合也表现出相似的功能结构[36, 52]。在跑步速度逐渐增加时（如 5、7、9、12 Km/h 的跑步机跑步场景），这 5 个时间固定协同的激活持续时长基本保持稳定，表明激活时长具有较强的速度不变性。然而，随着速度增大，各时间协同的激活曲线峰值出现向步态周期前部移动趋势[52]。表明，中枢神经系统可能通过调整激活峰值的时相来适应速度增高所带来的步态动力学需求。

## 4. 损伤与病理人群跑步中的肌肉协同模式

对损伤或病理状态下人群的肌肉协同模式进行研究，有助于深入理解慢性疾病或运动损伤对运动协调性的影响机制。已有文献表明，某些运动系统疾病会导致神经肌肉控制策略的合并或重组，进而影响动作执行效率与运动功能维持。

神经肌肉骨骼系统的结构与功能状态会显著影响协同控制策略。尽管目前病理性人群的相关研究主要以步行任务为主而，但其协同变化特征对理解跑步中的神经肌肉控制同样具有启示作用。已有证据表明，多种运动系统疾病会导致肌肉协同“合并”现象，即协同数量减少、协同结构趋于简单化，通常反映出神经控制能力下降。例如髌股疼痛综合征 (PFP) 患者[11]在步行中仅需要 3~4 个协同即可解释其肌电活动，而健康对照组则需要 4~5 个协同[11]。类似的，膝骨关节炎 (Knee Osteoarthritis, KOA)[9] 患者主要依赖 2~3 个协同，而健康年轻人则使用 3~4 个协同[9]。脑卒中患者的协同总数研究指出，偏瘫侧仅表现为 2~4 个协同，而非偏瘫侧则为 4~5 个[12]。

病理性人群还会通过调整协同中各肌肉的权重来应对身体的力学挑战。例如，重度 KOA 患者在支撑初期协同中表现为股二头肌与臀中肌权重明显上升[10]。在摆动初期协同中重度 KOA 患者的股四头肌外侧头权重显著小于较轻度患者，KOA 患者的胫骨前肌权重明显大于健康对照组，轻度 KOA 患者的腓肠肌外侧头权重显著小于健康个体[10]。此类变化可能代表患者在关节负荷过高或疼痛限制下形成的代偿策略，例如增加臀中肌和腘绳肌贡献以维持骨盆及关节稳定，或通过调整胫骨前肌与腓肠肌的力量平衡以缓冲地面冲击，降低膝关节瞬时载荷。

协同合并策略使原本应由相对独立协同控制的肌群在时间上趋于同步募集。例如，健康跑者在支撑相早期主要激活股四头肌，摆动相末期主要激活腘绳肌。然而，在 PFPS 患者中，常观察到协同合并现象，即出于稳定膝关节稳定的需要，股四头肌和腘绳肌在整个支撑相中持续共同激活[11]。这表明 PFP 患者倾向于采用复杂度更低、可控性更强但灵活性更差的神经控制信号，来补偿因疼痛或结构性改变引起的协调性下降。类似的合并在 KOA 患者中也广泛存在。KOA 步态中常见膝关节屈曲角增大与股四头肌-腘绳肌共收缩增强，二者均与协同结构的简化和功能模块的合并紧密相关，是 KOA 患者运动控制策略改变的典型制特征之一[9]。这种合并虽能够在短时间提高关节稳定性、减少疼痛引发的二次损伤风险，但同时也会降低运动效率，并增加代谢成本。

肌肉协同分析近年来被认为具有作为“运动控制障碍的神经标志物”的潜力[9, 11, 12]。这一量化指标能够反映神经肌肉系统的组织方式及其受损特征，可为康复训练提供重要参考。通过在训练中针对性地提升协同独立性、减少异常协同合并，有望改善神经控制能力与动作质量。例如，对长期存在股四头肌与臀肌异常共同激活的患者，可在康复过程中采用孤立强化训练、神经肌肉再教育或动作控制训练，恢复这些肌群的独立调控能力，从而减少不必要的共收缩，提高运动效率[11]。

## 5. 当前研究的局限性与未来的研究方向

尽管肌肉协同分析为理解跑步任务中的神经肌肉控制提供了重要的启发，但其方法与结论仍面临一系列局限。首先，协同模式的提取高度依赖表面肌电信号的质量，而肌电容易受电极放置、皮肤阻抗、运动伪影和肌电串扰等因素影响，使协同结构的稳定性受到干扰。其次，协同模式对预处理方法、归一化策略、协同分解算法、协同结构参数设置均高度敏感，不同研究之间在流程与标准上缺乏统一性，导致结果难以直接比较。此外，NMF 等线性分解方法本质上是统计建模工具，其提取的协同并不必然对应神经系统中的真实控制模块，因此需要警惕将数学分解结构完全等同于神经生理事实的风险。最后，协同分析常采用低维度近似来解释复杂的神经肌肉系统控制，但可能忽略肌肉调控的微小改变，如慢性损伤、疲劳、疼痛或补偿策略下的精细变化。

未来的肌肉协同研究方向可从以下几个方向进一步深化。(1) 制定任务特异性的标准化流程，如跑步研究的最低肌群覆盖要求、数据预处理标准化方案、协同提取的可靠方法及协同数量判定的多指标综合标准等；(2) 结合神经影像、脑脊髓控制机制与肌骨动力学建模，以阐明协同结构的神经机制与其生物力学约束的相互作用机制；(3) 结合深度学习算法等非线性模型，探索高维度的神经控制结构；(4) 通过训练干预、疲劳恢复、康复治疗或者疾病进展的纵向追踪，探索协同的可塑性与因果关系，并推动其在运动训练、人体科学及临床康复等领域的实践应用。

## 6. 结论

肌肉协同作为理解中枢神经系统如何在人体肌肉骨骼系统力学约束条件下实现高效控制的重要理论，已在跑步等动态运动中展现出独特价值。现有证据表明，成年人在不同速度、界面、足部着地模式等外部条件变化下，或在疲劳、老龄化等内部状态改变的影响下，协同数量保持相对稳定，而其空间肌肉权重与时间募集模式则表现出高度的可塑性。这种在协同框架内灵活调节时间与空间模式的特性反映了中枢神经系统在运动控制中对基本模块的稳定依赖以及对环境及生理变化的灵活适应。与此同时，病理性人群普遍存在协同合并与重组现象，协同模式有潜力成为"运动控制障碍的神经标志物"，为理解疾病机制与指康复训练提供新的视角。

尽管协同研究已取得显著进展，但在神经生理基础、实验标准化、方法可靠性、以及实践可推广性等方面存在诸多挑战。未来研究可结合神经影像、脑脊髓机制研究、肌骨动力学建模、深度学习算法及纵向研究，从数据驱动的"肌电协同"转变到真正的神经生理协同研究上。随着技术的发展和研究框架的完善，肌肉协同分析有望在运动科学、运动训练和康复医学实践中发挥更大的作用，为揭示运动控制机制和促进功能恢复提供坚实的理论基础与量化工具。

**利益冲突声明：**无

**作者贡献声明：**马晔负责论文框架设计、文献资料收集整理及论文初稿撰写与修改，林时鑫、傅升星、刘雨薇负责文献资料收集整理，郭辰仪负责论文修改，刘东威、侯美金负责论文框架设计、文献资料分析、论文撰写与修改。

**参考文献**